\documentclass{article}
\usepackage{a4,bbm,amsmath,amssymb,amsfonts,epsfig}

\newcommand{\wickn}[1]{\protect{:\!#1\!:}}

\newcommand{\rem}[1]{\vspace{1.5ex}{ 
{\bf Remark:} #1 }\vspace{1.5ex}}

\newcommand{\tsf}[2]{\protect{{\textstyle{\frac{ #1}{#2}}}}}

\newcommand{\msh}[1]{\protect{\frac{d{\bf #1}}{2\omega_{\bf #1}}}}

\newcommand{\omsh}[1]{\protect{\omega_{\bf #1}}}
\newcommand{\omshi}[3]{\protect{\omega_{{\bf
#1}_{#2}^{#3}}}}

\newcommand{\ra}[1]{\protect{{\bf #1}}}

\newcommand{\beqa}{\begin{eqnarray*}}
\newcommand{\eeqa}{\end{eqnarray*}}
\newcommand{\beqan}{\begin{eqnarray}}
\newcommand{\eeqan}{\end{eqnarray}}

\title{Ultraviolet Finiteness of the averaged Hamiltonian 
\\
on the noncommutative Minkowski space}

\author{{Dorothea~Bahns\thanks{currently at: 
Dipartimento di Matematica, Universit\`a di Roma ``La Sapienza'',
P.le Aldo Moro 2, 00185 Roma, Italy }}\\[2pt]
\small{II. Institut f\"ur Theoretische Physik}\\[-2pt]
\small{Universit\"at Hamburg} \\[-2pt]
\small{Luruper Chaussee 149}\\[-2pt]
\small{D-22761 Hamburg, Germany}\\[-2pt]
\small{bahns@mail.desy.de}}

\date{}

\begin{document}
\maketitle

\begin{abstract} 

\noindent It is shown that the Hamiltonian approach for a $\phi^3$-interaction 
on the 4-dimensional noncommutative Minkowski space leads to an ultraviolet finite 
$S$-matrix if the noncommutativity is averaged at 
each vertex.

\end{abstract}


Field theories on noncommutative spacetimes have recently received considerable
attention. For one thing, noncommutative spacetimes provide a 
covariant possibility to implement the idea that by Heisenberg's uncertainty
principle together with gravity measurements of arbitrarily small distances
should be impossible. On the other hand, field theories on such spacetimes
arise as low energy limits of string theories in background (magnetic) fields.

The present paper deals with an interaction Hamiltonian of a self-interacting
massive scalar field on the noncommutative Minkowski space as defined and
motivated in~\cite{dfr} which was given there as follows,
\begin{equation}\label{Hpos}
H_I(t)=\frac{g}{n!}\;
\mu \,\Big(\int_{q^0=t}\!\!\!d^3 q\;\,
\wickn{\phi^n (q)}\,\Big)\,.
\end{equation}
Here, the formal $3$-integral at time $t \in \mathbb R$ is defined as a
positive weight on the algebra $\mathcal E$ ``generated'' by quantum coordinates
$q^\mu$, $\mu=0,\dots,3$ (all technical
details are skipped here). The
corresponding generalized Weyl operators fulfill
\[
e^{ikq}e^{ipq}=e^{i(k+p)q}e^{-\frac i 2 kQ p}\,,\qquad kq=k_\mu q^\mu\,,
kQp=k_\mu Q^{\mu\nu}p_\nu\,,
\]
where the commutators $Q^{\mu\nu}=-i[q^\mu,q^\nu]$ are assumed to be central. 
The exponential $\exp({-\frac i 2 kQp})$
is referred to as the twisting. The joint spectrum $\Sigma$ of the $Q^{\mu\nu}$
is characterized by Poincar\'e invariant  conditions and turns out to be
homeomorphic to two copies of $TS^2$. In
particular, we have $\sigma^{0i}\neq 0$ for all $\sigma \in\Sigma$ (space-time
noncommutativity). The field $\phi(q)$ is defined using ordinary annihilation and creation
operators $a$ and~$a^*$,
\[
\phi(q)=\int dk\;  \check\phi(k)\otimes e^{ikq}
= (2\pi)^{-3/2}\int \msh k 
\;\big(\,a({ k})\,\otimes \,e^{-ikq}+a^*({k})\,\otimes
\,e^{ikq}\,\big)\,\big|_{k_0=\omsh k= \sqrt{\ra k^2 +m^2}}
\]
and is to be interpreted as a linear map from
states on $\mathcal E$ to smeared field operators, $
\omega\mapsto\phi(\omega)=\langle I\otimes \omega,\;\phi(q)\rangle=
\int dk\;\check \phi(k)\, \omega( e^{ikq}) 
$, which act on the ordinary
symmetric Fock space of the free scalar field. 
The double dots in~(\ref{Hpos}) indicate normal ordering in the ordinary sense,
applied to the annihilation and creation operators.
Now, the map $\mu$ is defined by $\mu (g(Q))=\int d\mu_\sigma
\,g(\sigma)$, where the measure $d\mu_\sigma$ is supported on $\Sigma_1$, a
subspace of $\Sigma$ which  is rotation and translation invariant, 
\begin{equation}\label{Sig1measure}
d\mu_\sigma=\,(8\pi)^{-1} \;d\vec e \,d \vec m \,\delta({\vec e\,}^2 - 1)\,
\delta^{(3)}(\vec e \mp \vec m)\qquad
\mbox{where } e_i=\sigma_{0i}\,, m_i=\epsilon_{ijk}\sigma_{jk}\,. 
\end{equation} 
The subspace $\Sigma_1$ arose in~\cite{dfr} in the discussion of states with minimal
uncertainties (best-localized states) and in the absence of a subspace of 
$\Sigma$ which would be invariant also under boosts {\em and} yield a  finite
measure, it was chosen for the definition of $H_I(t)$. The interpretation of
such integrations over parts of the spectrum $\Sigma$ is that already in the
definition of the interaction the effect of the noncommutativity is averaged in
a certain way -- contrary to an approach where $\sigma$ is mapped to a fixed
noncommutativity matrix $\theta$ serving as a fixed background. Note that
$H_I(t)$ can equivalently be written as an effective Hamiltonian on the
ordinary Minkowski space as follows  
\beqan\label{Hint}
H_I(t)\!\!&=& \!\!
\tsf{g}{n!}
\int_{\Sigma_{1}}\!\!d\mu_\sigma \int_{x^0=t}\!
\!\!\!d^3 {\bf x} \int\! \prod_{i\in N}dk_i 
\;\wickn{\check \phi(k_1) ...\, \check \phi(k_n)}
\,\exp\big({\textstyle 
-\tsf{i}{2}\sum\limits_{i<j} k_i \sigma k_j}\big)\,e^{-ix\,\sum k_i}\,,
\eeqan
where $N=\{1,\dots, n\}$. 
%
%
In~\cite{dfr} it was proposed to calculate expectation values from the
well-known formula 
$\sum_{r=0}^\infty \tsf{(-i)^r}{r!}\, 
\langle \Omega|\,T\,\Big(\,\phi_0(x_1)\dots \phi_0(x_n)\, \int dt_1 \dots dt_r \,
H_{I}(t_1)\dots H_{I}(t_r)\Big)\,|\Omega\rangle $, an approach which has later
been called the interaction point time-ordering approach. Equivalently, 
expectation values of the (formally) unitary $S$-matrix at $r$-th order of the
perturbative expansion in multi-particle states can be calculated as usual,
\beqan\label{expectS}
\nonumber
\langle p_1\dots {\rm in} \;| \,S_r \,| \;q_1 \dots {\rm in}\rangle
&=&{\textstyle \frac{(-i)^r}{r!}}\sum_{\pi\in P_r}
\int dt_1\dots dt_r\,\theta(t_{\pi_1}-t_{\pi_2})\dots
\theta(t_{\pi_{r-1}}-t_{\pi_r})\, \cdot
\\&&\qquad\qquad
\cdot\,\langle p_1\dots {\rm in} \;| \,\,H_I(t_{\pi_1})\dots H_I(t_{\pi_{r}})
 \,| \;q_1 \dots {\rm in}\rangle\,,\qquad\qquad
\eeqan 
where the improper energy-momentum eigenstates $| p_1\dots p_s
\rangle$ with the usual conventions are used. 

So far, different ways to derive graphical rules to simplify the calculation of
the $S$-matrix' expectation values on the noncommutative Minkowski space have
been investigated. For the purposes of this article, it is convenient to
use the rules in momentum space derived directly from (\ref{expectS})
in~\cite{bahnsdiss} which are reproduced in the appendix. Note that in the
derivation it is not necessary to specify the particular form of the nonlocal
kernel $\mathcal K(k_1,\dots,k_n)$ which in (\ref{Hint}) is given by the
integral of the twisting over $\Sigma_1$. In particular, it would also be
possible to use a point measure to map $\sigma$ to a fixed given
noncommutativity matrix $\theta$, in which case the rules indeed coincide with
those given in~\cite{sib}. The rules remain valid even if $\theta \notin\Sigma$.
The main step in the calculation is the evaluation
of the expectation values in the second line of (\ref{expectS}) by application
of the ordinary Wick reduction. For a kernel which is nonlocal in time, only
$3$-momenta are conserved at the vertex (though we still find overall
energy-momentum conservation), and in general no Feynman propagators appear,
see~\cite{BDFPuni} (and \cite{bahnsdiss} for details).


Before proceeding to the main subject of this article, a comment on the
infrared-ultraviolet mixing property seems appropriate. 
In the framework of the modified Feynman rules~\cite{filk}, the most serious
obstacle to applying ordinary renormalization procedures is a notorious mixing
of ultraviolet and infrared divergences which was first discussed
in~\cite{seib}. Roughly speaking, the problem is that  graphs which are
ultraviolet finite may develop hard infrared divergences when inserted into
larger graphs. It is by now established that the modified Feynman rules violate the optical
theorem  and that no Wick rotation seems to be available if they are applied in
a Euclidean setting, unless the time variable
commutes with the space variables or the special dilation
covariant noncommutativity from~\cite{dfr} (``lightlike
noncommutativity'') is employed. Notwithstanding, in order to illustrate the differences to
the Hamiltonian setting, two examples are discussed here. The first is the
following nonplanar contribution to the fish graph in Euclidean
$\phi^3$-theory, 
\[
\delta^{(4)}(q-q^\prime)\int dk dp \;\frac 1 {k^2+m^2} \frac 1
{p^2+m^2}\,e^{-ik\theta q}\,\delta^{(4)}(k+p-q)\,,
\]
and the second is given by the same expression with twisting $e^{-ik\theta p}$
instead of $e^{-ik\theta q}$. Here, $q$ and $q^\prime$ are the momenta which
enter and leave the diagram. These graphs (which are actually equal to one
another) are ultraviolet finite as long as $(\theta q)^2\neq 0$, but diverge
with leading order $\propto (\theta q)^{-2}$ if inserted into higher order
diagrams where $q$ may be zero.

Now consider these two graphs in a Hamiltonian setting with a Hamiltonian as 
given by (\ref{Hint}) but with fixed noncommutativity matrix
$\theta\in\Sigma$, i.e. $\mu(\sigma)=\theta$. Main
difference to the expressions discussed above is that now all momenta are on
the mass-shell. Choosing some time-ordering (which is irrelevant for the
argument), the Hamiltonian graph corresponding to the first contribution is
\[
 \delta^{(4)}(q-q^\prime)\int \msh k \msh p \;\frac 1 {-(\omsh k + \omsh p
-q_{0}) +i\epsilon} \,\delta^{(3)}(\ra k+\ra p-\ra q)\,e^{-ik\theta q}\,.
\]
Setting $q=(m,{\boldsymbol 0})$ and introducing spherical coordinates to
perform the $\ra k$-integration, where the $3$-axis is given by $-\ra e$ with
$e^i=\theta^{0i}$, we obtain 
\[
\int \msh k \frac 1{2\omega_{\ra k}} 
\frac 1 {-2\omsh k } \,e^{-ik_i\,\theta^{i0}\,m}
\propto
\int \!dr \;\frac{r}{|\vec e|\,m\,(r^2+m^2)^{3/2}}\,
 \;\sin({r\,|\vec e|\,m })\,.
\]
The integrand is well-defined for $r=0$, and for large $r$ it decreases as
fast as $1/r^2$, which is sufficient for the integral to remain well-defined.
Similarly, for the other contribution we find 
\[
 \delta^{(4)}(q-q^\prime)\int \msh k \msh p \;\frac 1 {-(\omsh k + \omsh p
-q_{0}) +i\epsilon} \,\delta^{(3)}(\ra k+\ra p-\ra q)\,e^{-ik\theta p}\,.
\]
Even for $q=0$ the above remains well-defined. To see this, introduce 
again spherical coordinates to obtain 
\begin{gather*}
\int \msh k \frac 1{2\omega_{\ra k}} 
\frac 1 {-2\omsh k } \,e^{-ik\theta (\omega_{\ra k},{-\ra k})}
\propto \int \!dr \;\frac{r}{|\vec e|\,(r^2+m^2)^2}\,
 \;\sin({+2\,r\,\sqrt{r^2+m^2}\,|\vec e| })\,.
\end{gather*}
Again, the integrand is well-defined in $r=0$, and for large $r$ decreases
like $1/r^3$ and hence fast enough for the integral to be
well-defined. Due to the fact that the energy is not  conserved at the vertex,
the two graphs now differ from one another.

These calculations are consistent with the fact that for commuting
time the modified Feynman rules and the Hamiltonian formalism coincide:
in the limit where $|\vec e|\rightarrow 0$ (commuting time) 
both integrands decrease like $1/r$ for ${ r\gg m^2}$, such that the 
integrals over $r$ diverge logarithmically as in the ordinary
case and coincide with the result of the corresponding calculation in the
setting of the modified Feynman rules. We have thus
seen that the situation in the Hamiltonian setting is more complicated than in
context of the modified Feynman rules, and that there is reason to hope that
this particular mixing of infrared and ultraviolet divergences is entirely
absent in the Hamiltonian approach with fixed noncommutativity matrix~$\theta$.
Similar results have been obtained in~\cite{fischer}. 


Let us now turn to the main subject of this article, namely the ultraviolet
behaviour in the Hamiltonian formalism if the integration over $\Sigma_1$ at
each vertex is performed (see also~\cite{bahnsdiss}). Explicitly, as already
shown in~\cite{dfr}, a direct calculation in spherical coordinates yields
\[
\int_{\Sigma_1} d\mu_\sigma\; e^{-\frac i 2 a\sigma b}=
\frac {\sin \gamma_+(a,b)}
{\gamma_+(a,b)} + \frac {\sin \gamma_-(a,b)}{\gamma_-(a,b)}\,,
\]
where $\gamma_\pm(a,b)= |- a_0 \ra b + b_0 \ra a \pm \ra a \times \ra b | $, 
and likewise, we find
\[
\int_{\Sigma_1} d\mu_\sigma\; e^{-\frac i 2 a\sigma b -\frac i 2 c\sigma f}=
\frac {\sin \beta_+(a,b,c,f)}
{\beta_+(a,b,c,f)} + \frac {\sin \beta_-(a,b,c,f)}{\beta_-(a,b,c,f)}\,,
\]
where
$\beta_\pm(a,b,c,f)= |- a_0 \ra b + b_0 \ra a \pm \ra a \times \ra b 
- c_0 \ra f + f_0 \ra c \pm \ra c \times \ra f | $. It follws that the averaged
Hamiltonian in $\phi^3$-theory takes the form
\beqa
H_{I}(t)&=&\tsf{g}{3!}
\int\! dk_1dk_2dk_3 \;
\Big(\,\frac {\sin \gamma_+}
{\gamma_+} + \frac {\sin \gamma_-}{\gamma_-}\Big)(k_1+k_2,k_2+k_3)
\\&&\qquad\qquad\qquad\quad
\;\wickn{\check \phi(k_1) \check \phi(k_2)\check \phi(k_3)}
\,\delta^{(3)}(\ra k_1+\ra k_2+\ra k_3)\,e^{+it\,\sum k_{i,0}}\,,
\phantom{\int}\eeqa
while for a $\phi^4$-self-interaction we have
\beqa
H_{I}(t)&=&\tsf{g}{4!}
\int\! dk_1...\,dk_4 \;
\Big(\,\frac {\sin \beta_+}{\beta_+} + \frac {\sin
\beta_-}{\beta_-}\Big)(k_1+k_2,k_2+k_3+k_4,k_3,k_4)
\\&&\qquad\qquad\qquad\quad\;\wickn{\check \phi(k_1) ...\, \check \phi(k_4)}
\,\delta^{(3)}(\ra k_1+...+\ra k_4)\,e^{+it\,\sum k_{i,0}}\,.
\phantom{\int}
\eeqa
As a matter of fact, it was already pointed out
in~\cite{filk} that differences to the case with fixed noncommutativity matrix
were to be expected, but they were not analysed. First of all, we
note that no cancellations of twistings from different vertices can occur. 
Moreover, the factors $1/\gamma_\pm$ and $1/\beta_\pm$ arising from the
integration over $\Sigma_1$ change the superficial degree of divergence, as
they both
decrease like an inverse square for large momenta. Under the assumption (which
will be weakened below) that the twisting is never trivial, we obtain the
following na\"\i ve  modified counting rules for a graph with $r$
vertices (in 4 dimensions):
\begin{itemize}
\vspace{-1.5ex}\item for each internal line: propagator $-1$, integration $+3$;
\vspace{-1.5ex}\item from momentum conservation: $-3$ at every vertex and once
$-1$ for energy conservation; 
\vspace{-1.5ex}\item from the energy factors: $-1$ for $r-1$ of the vertices;
\vspace{-1.5ex}\item from the integrated twisting: $-2$ at each one of the 
$r$ vertices.
\end{itemize}
\vspace{-1ex}
Therefore, for a theory with $\phi^n$-self-interaction the superficial degree
of divergence is  
\[
\omega(\Gamma)= 2b-6r+4= n\cdot r-e - 6r
+4=(n-6)r-e+4=\omega_{ord}(\Gamma)-2r\,,
\] 
where $e$ is the number of external lines, $b$ the number of internal lines,
i.e. $2b=n\cdot r - e$, and where $\omega_{ord}$ denotes the ordinary degree of
divergence. As usual, $\omega < 0 $ implies that a graph is superficially
finite. Note that the oscillating sine-factors which may further improve the
ultraviolet behaviour have not been taken into account here.

\rem{Application of the counting rules to $\phi^3$-theory yields 
ultraviolet-finiteness for all graphs apart from the tadpole 
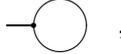
\[
\begin{picture}(40,13)
\put(0,3){\line(3,0){10}}
\put(10,3){\circle*{2}}
\put(20,3){\circle{20}}
\end{picture}\,,
\]
since for $e=1$, $\omega$ will be less than zero for any graph with
$r>1$,  and for $e> 1$, $\omega$ is always less than zero  (usually, $\omega<0$
for $r>4-e$, $e\geq 1$). }

Let us convince ourselves of the finiteness of the fish graph in an explicit
calculation. To do so, we apply the rules from the appendix as they
stand, with the only difference that the integration over $\Sigma_1$ is
performed. If it is true that the twisting is never trivial, it suffices for
our purposes to pick one  particular contribution (instead of using the
symmetrized kernels $\mathfrak S$). Consider, for instance, the contribution
where the second and the third field of the earlier vertex are contracted with
the second and the third field of the later vertex, and $q$ is an external
momentum leaving  the later vertex, while $q^\prime$ enters at the earlier one.
Then the rules yield
\begin{gather*}
 \delta^{(4)}(q-q^\prime)\int \msh k \msh p \;\frac {\delta^{(3)}(\ra
k+\ra p-\ra q)} {-(\omsh k + \omsh p -q_{0}) +i\epsilon} \,
\Big(\Big(\,\frac {\sin \gamma_+}
{\gamma_+} + \frac {\sin \gamma_-}{\gamma_-}\Big)(-q+k,k+p)\Big)^2
\\= \delta^{(4)}(q-q^\prime)\int \msh k \frac1{2 \omega_{\ra q-\ra k} }
\;\frac {1} 
{-(\omsh k + \omega_{\ra q-\ra k} -q_{0}) +i\epsilon} \,
\Big(\,2\,\frac {\sin \tilde\gamma_{\;\ra k,\ra q}}
{\tilde\gamma_{\;\ra k,\ra q}}\Big)^2\,,
\end{gather*}
where 
\[
\tilde\gamma_{\;\ra k,\ra q}=
\gamma_\pm(-q+k, (\omsh k + \omega_{\ra q-\ra k},{\bf 0})
=(\omsh k + \omega_{\ra q-\ra k})\,|\ra k-\ra q| \,,
\]
such that, after introducing spherical coordinates, we see that the integrand
decreases as fast as $|\ra k|^{-5}$ (i.e. $\omega=-4$), rendering a
well-defined integral. Note that this remains true for $\ra q=0$. It is
interesting to observe that the sine-factor is absent in part of the above
expression due to the fact that it appears in a square.

In the same manner, application of the na\"\i ve counting rules  yields
finiteness of $\phi^4$-theory.  The main difference compared to the ordinary
case is that $\omega$ is no longer independent of the number of vertices. As
usual, there are no graphs with an odd number of external legs $e$ (recall that
the number of contractions in a graph arising in $\phi^n$-theory is $\frac 1 2
(r\cdot n - e)$ such that for $n=4$, $4\,r - e$ must be even). Hence, the
smallest (non-zero) possible number of external legs is $2$.

\rem{ Provided that the twisting is never trivial, the superficial degree
of divergence~$\omega$ for graphs in $\phi^4$-theory with at least two external
legs,  $e\geq 2$, (i.e. all non-vacuum graphs), is less than $0$ for $r\geq 2$
(in the ordinary case $r\geq 4$ is necessary). By the same argument, graphs
with $e\geq 4$ are also superficially finite, as in this case $\omega\leq-2r$ 
(usually, $ \omega$ is independent of~$r$, $\omega=-e+4$ such that the number
of external legs must be greater than $6$ in order to find superficially finite
graphs). It thus follows from the na\"\i ve counting rules that all graphs
apart from vacuum graphs and graphs containing tadpoles (where $e=2$, $r=1$ is
possible) are superficially finite. }

However, things are not quite so simple. The reason is that there are some
graphs in which the twisting at a vertex may become trivial due to particular
relations between the momenta entering and leaving it. Before investigating
this question, we recall that  $\Sigma_1$ is the orbit of $\sigma^{(0)}$ (which
is non-degenerate) under the action of the orthogonal group, and, hence, that
the space where the twisting becomes trivial is the same as the one where
$\gamma_\pm=0$ or $\beta_\pm=0$, respectively.

In $\phi^3$-theory the situation is still fairly simple. At the vertex, we make
use of $3$-momentum conservation to replace $\ra k_3$, which for the twisting
yields ($\mu_i 
=\pm\, \omshi k i {}$, $\mu_{i+j}
=\pm\, \omega_{\ra k_i +\ra k_j}$),
\[
-(\mu_1+2\mu_2+\mu_{1+2})\,\sigma^{0i}\,k_{1,i}
+(\mu_2+\mu_{1+2})\,\sigma^{0i}\,k_{2,i}+k_{1,i}\,\sigma^{ij}\,k_{2,j}\,.
\]
It follows that the only dependence between $k_1$ and $k_2$ which can arise
from the graph theory and renders the twisting trivial is that $k_1=-k_2$,
including the energy-component (it follows that $\ra k_3=0$). This is only
possible if  either: $k_1$ and $k_2$ are the only external momenta of the
graph, and $k_3$ connects the vertex under consideration with a bubble graph
(or a number of bubble graphs),
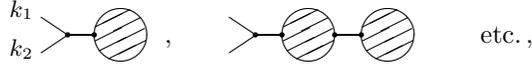
\[
\qquad\begin{picture}(35,13)
\put(0,3){\line(-3,2){10}}
\put(0,3){\line(-3,-2){10}}
\put(-22,10){\small $k_1$}
\put(-22,-5){\small $k_2$}
\put(0,3){\circle*{2}}
\put(0,3){\line(3,0){10}}
\put(10,3){\circle*{2}}
\put(12,-3){\line(2,1){17}}
\put(15.5,-6){\line(2,1){14}}
\put(10.5,1){\line(2,1){17}}
\put(10.5,6){\line(2,1){13}}
\put(20,3){\circle{20}}
\end{picture}\,,
\qquad\mbox{}\quad
\begin{picture}(65,13)
\put(0,3){\line(-3,2){10}}
\put(0,3){\line(-3,-2){10}}
\put(0,3){\circle*{2}}
\put(0,3){\line(3,0){10}}
\put(10,3){\circle*{2}}
\put(12,-3){\line(2,1){17}}
\put(15.5,-6){\line(2,1){14}}
\put(10.5,1){\line(2,1){17}}
\put(10.5,6){\line(2,1){13}}
\put(20,3){\circle{20}}
\put(30,3){\circle*{2}}
\put(30,3){\line(3,0){10}}
\put(40,3){\circle*{2}}
\put(42,-3){\line(2,1){17}}
\put(45.5,-6){\line(2,1){14}}
\put(40.5,1){\line(2,1){17}}
\put(40.5,6){\line(2,1){13}}
\put(50,3){\circle{20}}
\end{picture}
\qquad\mbox{etc.}\,,
\]
or: $k_1$ and $k_2$ are internal momenta of a bubble, one of which enters
and one of which leaves the only vertex of the bubble where some
(external or internal) momentum ($k_3$) enters, 
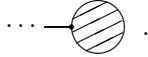
\[
\begin{picture}(45,13)
\put(-5,3){\dots}
\put(10,3){\line(3,0){10}}
\put(20,3){\circle*{2}}
\put(22,-3){\line(2,1){17}}
\put(25.5,-6){\line(2,1){14}}
\put(20.5,1){\line(2,1){17}}
\put(20.5,6){\line(2,1){13}}
\put(30,3){\circle{20}}
\end{picture}\,.
\]
We conclude in particular, that in the fish graph the twisting is never
trivial. We know already that the tadpole graph has to be subtracted, which
may be taken care of by using a normally ordered interaction term. Therefore,
the important question is, whether the only other graph which has to be
renormalized in ordinary $\phi^3$-theory,
\begin{picture}(35,9)(0,2.5)
\put(0,3){\line(3,0){10}}
\put(10,3){\circle*{2}}
\put(20,-7){\circle*{2}}
\put(20,13){\circle*{2}}
\put(20,-7){\line(0,3){20}}
\put(20,3){\circle{20}}
\end{picture},
is finite or not. If it is, $\phi^3$-theory is ultraviolet finite in this
approach. 

Taking into account that one of the twistings may become trivial in this graph,
the counting rules still yield the desired result:  since the graph ordinarily
diverges logarithmically, the remaining two vertices where the twisting stays
non-trivial suffice to render a well-defined integration ($\omega=-4$). This
can be confirmed by a tedious explicit calculation.

In $\phi^4$-theory the situation is more complicated. Here, $3$-momentum at the
vertex yields (with $\mu_i =\pm\, \omshi k i {}$, $\mu_{i+j+l}
=\pm\, \omega_{\ra k_i +\ra k_j+\ra k_l}$),
\begin{gather*}
-(\mu_1+2\mu_2+2\mu_3+\mu_{1+2+3})\,\sigma^{0i}\, k_{1,i}
-(\mu_2+2\mu_3+\mu_{1+2+3})\,\sigma^{0i}\, k_{2,i}
\\-(\mu_3+\mu_{1+2+3})\,\sigma^{0i}\, k_{3,i}
+k_{1,i}\,\sigma^{ij}\, (k_{2,j}+k_{3,j})+k_{2,i}\,\sigma^{ij}\, k_{3,j}\,.
\end{gather*}
Conjecture: The only dependence between the momenta which can arise in the graph
theory and which results in the above being 0, is: $k_1=-k_2$ and the
energy-components of $k_3$ and $k_4$ have opposite sign ($\ra k_3=-\ra k_4$
follows from $\ra k_1=-\ra k_2$).

The only non-vacuum graph in which a vertex with such a dependence arises is
the following:
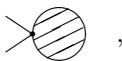
\[
\begin{picture}(40,13)
\put(10,3){\line(-3,2){10}}
\put(10,3){\line(-3,-2){10}}
\put(10,3){\circle*{2}}
\put(20,3){\circle{20}}
\put(12,-3){\line(2,1){17}}
\put(15.5,-6){\line(2,1){14}}
\put(10.5,1){\line(2,1){17}}
\put(10.5,6){\line(2,1){13}}
\end{picture}\,,
\]
where one of the internal momenta of the bubble enters, the other leaves the
vertex, and where $4$-momentum conservation of the external momenta ensures that
their energy-components have opposite sign. Again, by application of the
counting rules, we conclude that, even taking into account that one of the
twistings may become trivial, only the tadpole graph 
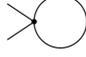
\[
\begin{picture}(40,13)
\put(10,3){\line(-3,2){10}}
\put(10,3){\line(-3,-2){10}}
\put(10,3){\circle*{2}}
\put(20,3){\circle{20}}
\end{picture}\,\,
\]
diverges, since in all other cases, $\omega$ is at least equal to $-2$.
Therefore, all possible graphs apart from vacuum graphs (and graphs containing
tadpoles) are superficially finite:
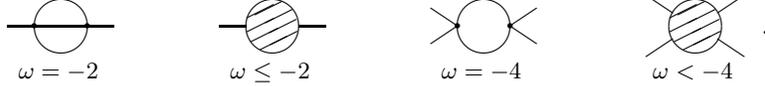
\[
\begin{picture}(40,23)(0,-10)
\put(0,3){\line(3,0){40}}
\put(10,3){\circle*{2}}
\put(30,3){\circle*{2}}
\put(20,3){\circle{20}}
\put(4,-17){{\small $\omega=-2$}}
\end{picture}
\qquad\qquad
\begin{picture}(40,23)(0,-10)
\put(0,3){\line(3,0){10}}
\put(30,3){\line(3,0){10}}
\put(20,3){\circle{20}}
\put(12,-3){\line(2,1){17}}
\put(15.5,-6){\line(2,1){14}}
\put(10.5,1){\line(2,1){17}}
\put(10.5,6){\line(2,1){13}}
\put(4,-17){{\small $\omega\leq-2$}}
\end{picture}
\qquad\qquad\begin{picture}(40,23)(0,-10)
\put(30,3){\line(3,2){10}}
\put(30,3){\line(3,-2){10}}
\put(10,3){\line(-3,2){10}}
\put(10,3){\line(-3,-2){10}}
\put(10,3){\circle*{2}}
\put(30,3){\circle*{2}}
\put(20,3){\circle{20}}
\put(4,-17){{\small $\omega=-4$}}
\end{picture}
\qquad\qquad
\begin{picture}(40,23)(0,-10)
\put(28.5,8){\line(3,2){10}}
\put(28.5,-2){\line(3,-2){10}}
\put(11.5,8){\line(-3,2){10}}
\put(11.5,-2){\line(-3,-2){10}}
\put(20,3){\circle{20}}
\put(12,-3){\line(2,1){17}}
\put(15.5,-6){\line(2,1){14}}
\put(10.5,1){\line(2,1){17}}
\put(10.5,6){\line(2,1){13}}
\put(4,-17){{\small $\omega<-4$}}
\put(45,0){.}
\end{picture}
\]

Finally note that in an approach based on the averaged Hamiltonian, no
infrared-ultraviolet mixing in the sense discussed above appears at all.


\vspace{3ex}
{\bf Acknowledgement} 

I would like to thank K. Fredenhagen, S. Doplicher and G. Piacitelli for many 
fruitful discussions on field theories on noncommutative spacetimes.


\begin{appendix}

\section{Rules}

The following rules which were derived in~\cite{bahnsdiss} are valid for
Hamiltonians with arbitrary nonlocal kernels. To simplify the combinatorics,
such kernels should be symmetrized in order to make the momenta of one vertex
indistinguishable. For the Hamiltonian (\ref{Hint}) 
this means that the customary symmetrized vertex factors 
\[
\mathfrak S(\sigma; k_1,\dots,k_n)\stackrel{\rm def}{=}
\tsf{1}{n!}\sum_{\pi \in P_n}
\exp\big(-{\textstyle\frac{i}{2} \sum\limits_{i<j}} 
k_{\pi_i} \sigma k_{\pi_j}\big)
\]
are employed. It is,  however, also possible to consider all contributions
separately.
Note that the rules are stated for arbitrary measures on $\Sigma$ (in
fact, even the map $\sigma\mapsto \theta$, where $\theta \notin \Sigma$ is
admitted), and the appropriate operation is to be performed as a last step. 

\begin{enumerate}
\item Draw all ordinary connected Feynman graphs of the process under
consideration, characterized by the number of vertices ($r$ vertices at $r$-th
order of the perturbative expansion) and the number of external momenta. Since
we consider a normally ordered interaction term, no tadpole  graphs need to be
considered, such that no internal line starts and ends at the same vertex.
Consider all possibilities to distribute the external momenta to the vertices
separately. 

\item Pick one of the above graphs and assign times
$t_{1},\dots, t_{r}$ to its vertices. Choose a particular time-ordering.

\item For every internal line write down a mass-shell integral 
$
\frac1 {(2\pi)^{3}}\int \msh k 
$,
where $k=(\omsh k,\ra k)$ labels the directed line connecting the earlier
vertex with the later one. $k$ thus flows out of the earlier vertex into the
later one.

\item For each vertex $j$ apart from the earliest one 
write down the following energy factors,
$
\frac{i}{2\pi} ({-\sum_{i} k_{i,0} +i\epsilon})^{-1}$,
where the sum runs over the $0$-components of the following momenta: 
\begin{itemize}
\item the internal momenta flowing into the vertex (with a
$+$ sign), i.e. the ones labelling inner lines starting at earlier vertices and
ending in the vertex under consideration;
\item the internal momenta flowing into any of the later vertices (with a
$+$ sign), provided they
start at earlier vertices than the vertex under consideration;

\item the external momenta which flow into or out of the
vertex under consideration or any of the later vertices of the
graph. 
Here, outgoing momenta enter with a $-$ sign, while the ones flowing into a
vertex enter with a $+$ sign. 
\end{itemize}

\item At each vertex impose $3$-momentum conservation 
$(2\pi)^{3}\delta^{(3)}(...)$, where incoming momenta enter with a $+$ sign,
outgoing momenta with a $-$ sign.  Impose overall energy conservation of the
external momenta $2\pi\,\delta(...)$, where incoming momenta enter
with a $+$ sign, momenta flowing out of  the graph with a $-$ sign. 
For every external momentum multiply with a factor $(2\pi)^{-3/2}$. 
\item 
At each vertex, the twisting $\mathfrak S$ is determined
by the following rules:
\begin{itemize}
\item an external momentum leaving the vertex enters with a $-$
sign; 
\item an external momentum flowing into the vertex enters with a
$+$ sign; 
\item an internal momentum enters with a $+$ sign, if the vertex is the endpoint
of the momentum's line, and it enters with a $-$ sign if the vertex is the
starting point of the momentum's line. 
\end{itemize}
As the twistings $\mathfrak S$ are symmetrized, the order of the momenta in
$\mathfrak S$ is arbitrary. 

\item 
For each of the $r$ vertices  multiply with a factor $\tsf{g}{n!}$. Calculate
the symmetry factor of the diagram as usual and multiply the expression with 
it. Multiply with a factor~$\tsf{(-i)^r}{r!}$.
\end{enumerate} 

If $\mathfrak S$ is independent of the energy components, the graphs can be
added in such a manner as to yield Feynman propagators, see~\cite{bahnsdiss}.

\end{appendix}



\end{document}